\documentclass[pdflatex,sn-mathphys-num]{sn-jnl}

\usepackage{epsfig}
\usepackage{subcaption}
\usepackage{calc}
\usepackage{amssymb}
\usepackage{amstext}
\usepackage{amsmath}
\usepackage{amsthm}
\usepackage{multicol} 
\usepackage{url}
\usepackage[most]{tcolorbox} 
\usepackage{booktabs}
\usepackage{multirow}
\usepackage{tabularx}
\usepackage{geometry}

\usepackage[table]{xcolor}
\definecolor{maroon}{cmyk}{0,0.87,0.68,0.32}

\usepackage{amssymb}
\usepackage[figuresright]{rotating}

\usepackage{listings} 

\definecolor{codegreen}{rgb}{0,0.6,0}
\definecolor{codegray}{rgb}{0.5,0.5,0.5}
\definecolor{codepurple}{rgb}{0.58,0,0.82}
\definecolor{backcolour}{rgb}{0.95,0.95,0.92}

\lstdefinestyle{mystyle}{
    backgroundcolor=\color{backcolour},   
    commentstyle=\color{codegreen},
    keywordstyle=\color{magenta},
    numberstyle=\tiny\color{codegray},
    stringstyle=\color{blue},
    basicstyle=\ttfamily\footnotesize,
    breakatwhitespace=false,         
    breaklines=true,                 
    captionpos=b,                    
    keepspaces=true,                 
    numbers=left,                    
    numbersep=5pt,                  
    showspaces=false,                
    showstringspaces=false,
    showtabs=false,                  
    tabsize=2
}

\begin{document}

\definecolor{main}{HTML}{5989cf}    % setting main color to be used
\definecolor{sub}{HTML}{cde4ff}     % setting sub color to be used
\newtcolorbox{boxE}{
    enhanced, % for a fancier setting,
    boxrule = 0pt, % clearing the default rule
    borderline = {0.75pt}{0pt}{main}, % outer line
    borderline = {0.75pt}{2pt}{sub} % inner line
}

\title{Towards Analysing Invoices and Receipts with Amazon Textract}

\author{Sneha Oommen, Gabby Sanchez, Cassandra T. Britto, Di Wang, Jordan Chiou, Maria Spichkova}

\affil{ \orgname{RMIT University}, \orgaddress{ \city{Melbourne}, \country{Australia}}}

\abstract{
This paper presents an evaluation of the AWS Textract in the context of extracting data from receipts.
We analyse Textract functionalities using a dataset that includes receipts of varied formats and conditions. 
Our analysis provided a qualitative view of Textract’s strengths and limitations. 
While the receipts' totals were consistently detected, we also observed typical issues and irregularities that were often influenced by image quality and layout. Based on the analysis of the observations, we propose mitigation strategies.  
}
  
\maketitle
%========================================
\section{Introduction}

Optical Character Recognition (OCR) techniques~\cite{chaudhuri2016optical} have been utilised in various contexts to recognise text from provided images.   
For example, the OCR approaches have been introduced to manage handwritten documents, see, e.g., the systematic literature review by Memon et al.~\cite{memon2020handwritten}. 
The idea of OCR was patented almost 100 years ago, in 1929 in Germany and in 1933 in the U.S., see the historical review conducted by Mori et al.~\cite{mori1992historical} for more details. 
Since then, the OCR technology has evolved significantly. The increase in accuracy and overall performance has been especially remarkable over the last decade.
Numerous new approaches have emerged, making OCR a viable option for analysing images and scanned documents related to everyday tasks. 

Our study focused on AWS Textract’s accuracy in extracting data from receipts. 
AWS Textract is a managed document analysis service that uses machine learning to extract text and
key-value pairs from scanned documents~\cite{AWSTextract1}. For expense data, the AnalyzeExpense API provides a pre-trained model optimised for receipts and invoices~\cite{AWSTextract2}.

\emph{Contributions:}  
We analyse Textract functionalities using a dataset that includes receipts of varied formats and conditions. 
Our analysis provided a qualitative view of Textract’s strengths and limitations.  
We analysed Textract’s performance across multiple receipt formats and vendors, focusing on its field detection accuracy, consistency, and completeness. These findings were used to highlight strengths, limitations, and areas for potential improvement before integrating Textract into the end-to-end ReceiptCAT workflow. 
To evaluate this extraction stage, we assessed AWS Textract’s AnalyzeExpense API, which converts
receipt images into structured JSON output. The goal was to determine how effectively Textract could detect and extract the essential expense fields across a diverse sample of Australian receipts and to identify recurring extraction errors or inconsistencies. 
While the receipts' totals were consistently detected, we also observed typical issues and irregularities that were often influenced by image quality and layout. Based on the analysis of the observations, we propose mitigation strategies.

%========================================
\section{Related Work}
\label{sec:related}

In our previous studies~\cite{spichkova2019easy,spichkova2019comparison}, we investigated the performance of Google Cloud Vision and AWS Rekognition for the automation of the meter reading process for the standard (non-smart) meters using computer vision techniques. The study demonstrated that AWS Rekognition provides better results for reading of the data from non-smart meters. However, over the last years, many other studies have been conducted based on AWS technologies. In what follows, we provide a brief overview of them.
 
\subsection{AWS Textract}

Hegghammer~\cite{hegghammer2022ocr} conducted an experiment comparing the OCR performance of  Tesseract, Amazon Textract, and Google Document AI. The study demonstrated that Document AI and Textract performed significantly better than Google Document AI. 
A comparative analysis of Tesseract and Textract has been conducted by Modi et al.~\cite{modi2024optical}. The authors concluded that Textract demonstrates better accuracy and scalability.  
Padmaja et al.~\cite{padmaja2024comparative} aimed to compare the performance of Textract and Tesseract OCR in recognition of handwritten text from images. Their study also confirmed that Textract provides higher accuracy of text recognition. 
Our study, in contrast, focuses on AWS Textract, as this solution has demonstrated high performance in other studies. 

Deborah et al.~\cite{deborah2024efficient} analysed AWS Textract services, but haven't provided a detailed evaluation of the results and potential limitations. 
Kanani et al.~\cite{kanani2024information} conducted a study on data extraction from medical reports using AWS Textract, but also did not present any evaluation or discussion of potential limitations. 
Kumar et al.~\cite{kumar2023comparative} also conducted a study on data extraction from medical reports, where the authors used both handwritten and printed records. The authors compared the performance of AWS Textract, Google Vision, and PyTesseract.  

Kodali et al.~\cite{kodali2023automated} presented a Textract-based plagiarism detection system. The system was developed for handwritten text in English, but the evaluation wasn't presented by the authors. 

In contrast to the above works, our study aims to investigate the limitations of AWS Textract and to propose corresponding mitigation strategies.

 Madake and Pandey~\cite{madake2023tabular} worked on data extraction from tables using  Google’s vision API as well as other approaches, but their solution requires too many manual steps and is hardly applicable to extracting data from shopping receipts.

Lin et al.~\cite{lin2025visual} proposed a tool for data extraction from template-based documents. The authors claim that their tool outperforms Textract, Evaporate,  and Azure Document Intelligence. In contrast to this work, we aim to evaluate Textract, focusing on a particular application domain, shopping receipts, where the format of the templates might be very diverse.

\subsection{Research embedded in teaching and learning process}

This study has been conducted within the employability initiative 
proposed at the RMIT University, see \cite{spichkova2017autonomous,simic2016enhancing}. 
Enhancing students' exposure to research activities is a crucial part of the curriculum. Nevertheless, there is no agreement on the best way to achieve this.. 
Holz at al.~\cite{holz2006research} proposed a framework for teaching Computing
Research Methods (CRM). 
Weinman et al.~\cite{weinman2015teaching} proposed to offer summer undergraduate research experiences for students. 
Fernandez~\cite{fernandez2024generalizing} presented the initiative introduced in the University of Texas to provide course-based research experience for undergraduate students, where students work on research tasks individually. 
Host~\cite{host2002introducing} proposed to include empirical software engineering methods in the study curriculum within project-based courses, which is especially close to the spirit of our initiative.

To foster the curiosity of Bachelor's and Master's students regarding research in Software Engineering, we propose incorporating research and analysis components into projects as an optional bonus task. The research projects were typically short, corresponding to approx. one week of a full-time study. They have been typically conducted by teams of four or five students, sponsored by industrial partners. The topic of the research projects aligned with the topics related to the software development project to be conducted within the semester, which led to a broad range of topics to be covered by several teams.
For example, Sun et al.~\cite{sun2018software} and Clunne-Kiely et al.~\cite{clunne2017modelling} analysed Human-Computer interaction aspects related to an autonomous humanoid robot, whose task is to guide visitors through a VXlab. 
Spichkova et al.~\cite{spichkova2020gosecure}  presented an automated solution for scanning security vulnerabilities in Google Cloud Platform projects, addressing gaps in the capabilities of existing solutions that detect common security issues. 
George et al.~\cite{george2020usage} conducted a study on the visualisation of the usage of the AWS service. 
Zhao et al.~\cite{zhao2025visualisation} also worked on the visualisation aspects, but their study focused on the CIS benchmark scanning results. 
Christianto et al.~\cite{christianto2017software} also worked on visualisation, but in a different context: the study focused on the visualisation and analysis of data collected from vertical transport facilities.

% %========================================
\section{{Case study: Methods}}
\label{sec:methodology}

The evaluation used 118 receipts collected from Australian vendors. The dataset was manually gathered
by the ReceiptCAT development team, consisting of five members. A total of 77 receipts (65\%) were
physical receipts photographed using three different smartphones: Samsung Galaxy A52 (Android 14),
Google Pixel 7 Pro (Android 16), and iPhone 13 Pro (iOS 18.6.2). The remaining 41 receipts (35%)
were electronic receipts obtained as screenshots of digital copies. 

The study was limited to receipts in JPG and PNG formats, taken with a camera or via screenshots from mobile phone apps (both Android and iOS). Thus, all 118 receipts were stored as image files, and no PDF-based receipts were included.

Due to time constraints, no image stratification was applied during data collection. We did not modify or standardise the photographs to control for factors such as lighting, clarity, or angle.
As a result, the dataset contains a natural mix of image conditions rather than predefined experimental categories. This approach allowed the team to focus on testing Textract’s performance on realistic user-generated receipts, with controlled analysis of image quality and receipt condition identified as a future evaluation task.
The characteristics of the collected receipts, including the proportion of physical and electronic copies, the devices used to capture the photos, and the observed physical conditions of the receipts, are summarised in Tables~\ref{tab:data-format} and \ref{tab:receipt_characteristics}.

\begin{table}[ht!]
\centering
\caption{Dataset used in the case study: Receipt format}
\label{tab:data-format}
\label{tab:receipt_characteristics}
\renewcommand{\arraystretch}{1.3}
\begin{tabularx}{\textwidth}{p{3cm} p{4cm} c c}
\toprule
 & \textbf{Category} & \textbf{Receipt Count} & \textbf{Percentage (\%)} \\
\midrule
\multirow{2}{*}{\textbf{Receipt Format}}
 & Physical & 77 & 65.25 \\
 & Electronic (Screenshots) & 41 & 34.74 \\
\bottomrule
\end{tabularx}
\end{table} 

\begin{table}[ht!]
\centering
\caption{Dataset used in the case study: Physical receipts}
\label{tab:receipt_characteristics}
\renewcommand{\arraystretch}{1.3}
\begin{tabularx}{\textwidth}{p{4.5cm} p{2.5cm} c c}
\toprule
\textbf{Attribute} & \textbf{Category} & \textbf{Receipt Count} & \textbf{Percentage (\%)} \\
\midrule
\multirow{3}{*}{\textbf{Phone Model}}
 & Samsung Galaxy A52 (Android 14) & 33 & 27.97 \\
 & Google Pixel 7 Pro (Android 16) & 26 & 22.03 \\
 & iPhone 13 Pro (iOS 18.6.2) & 18 & 15.26 \\
\midrule
\multirow{2}{*}{\textbf{Image quality}}
 & Clear & 62 & 52.54 \\
 & Blurred & 15 & 12.71 \\
\midrule
\multirow{3}{*}{\textbf{Receipt condition}}
 & Folded & 35 & 29.66 \\
 & Crumpled & 24 & 20.34 \\
 & Flat & 18 & 15.25 \\
\midrule
\multirow{2}{*}{\textbf{Angle of receipt in  the image}}
 & Levelled & 25 & 21.19 \\
 & Skewed & 52 & 44.07 \\
\bottomrule
\end{tabularx}
\end{table}

%--------------------------------------------
\subsection{Dataset}
The dataset represents 59 distinct vendors, with receipts sourced from a mix of large retail chains, local stores, and service-oriented businesses. 
This composition reflects the typical spending patterns, with major retailers accounting for approximately one-third of the dataset, and the remainder comprising smaller retail and service vendors. The largest contributors were Woolworths and Woolworths Metro (one of the largest supermarket chains in Australia), which together accounted for 24\% of all receipts. 
Most other vendors were presented by only a few receipts each, spanning industries such as groceries, dining, beauty, electronics, pharmacy, utilities, and professional services. The distribution of receipts across vendors with more than 1 receipt is presented in Table~\ref{tab:vendor-receipts}.

\begin{table}[ht!]
\centering
\caption{Receipt distribution by vendor}
\label{tab:vendor-receipts}
\renewcommand{\arraystretch}{1.3}
\begin{tabular}{lcc}
\toprule
\textbf{Vendor} & \textbf{Receipt Count} & \textbf{Percentage (\%)} \\
\midrule
Woolworths            & 19 & 16.10 \\
Woolworths Metro      & 10 & 8.47  \\
Coles                 & 7  & 5.93  \\
Medical Hub RMIT      & 6  & 5.08  \\
BIGW                  & 5  & 4.24  \\
Chemist Warehouse     & 4  & 3.39  \\
Sephora               & 4  & 3.39  \\
IGA                   & 4  & 3.39  \\
Aldi                  & 4  & 3.39  \\
Halaya                & 3  & 2.54  \\
Aussie Broadband      & 3  & 2.54  \\
Joy Market            & 2  & 1.69  \\
\bottomrule
\end{tabular}
\end{table}

\begin{table}[ht!]
\centering
\caption{Receipt distribution by vendor type}
\label{tab:vendor-type}
\renewcommand{\arraystretch}{1.2}
\begin{tabular}{lcc}
\toprule
\textbf{Vendor Type} & \textbf{Receipt Count} & \textbf{Percentage (\%)} \\
\midrule
Groceries                   & 47 & 39.83 \\
Eating Out \& Takeaway      & 30 & 25.42 \\
Professional Services       & 7  & 5.93  \\
Personal Care \& Beauty     & 6  & 5.08  \\
Department Store            & 5  & 4.24  \\
Pharmacy                    & 5  & 4.24  \\
Utilities \& Bills          & 5  & 4.24  \\
Stationery \& Office        & 3  & 2.54  \\
Clothing \& Footwear        & 2  & 1.69  \\
Electronics \& Tech         & 2  & 1.69  \\
Gifts \& Occasions          & 2  & 1.69  \\
Home \& Cleaning            & 2  & 1.69  \\
Sports \& Fitness           & 1  & 0.85  \\
Travel \& Holidays          & 1  & 0.85  \\
\bottomrule
\end{tabular}
\end{table}

To better understand the diversity of receipt sources, all vendors were classified into predefined business categories. The breakdown of receipts across these categories is shown in Table~\ref{tab:vendor-type}.  
The largest vendor type in the dataset was \emph{Groceries}, representing 40\% of all receipts, followed by \emph{Eating Out \& Takeaway} at 25\%. 
Together, these two categories accounted for approximately two-thirds of all receipts collected. This reflects typical household spending, excluding the categories of rent/mortgage.

%--------------------------------------------
\subsection{Data extraction methods}

Each receipt was uploaded to an Amazon S3 bucket configured with an event notification that invoked the receipt extraction Lambda function. 
The lambda called \texttt{AnalyzeExpense API} (AWS Textract) for each new object, parsed the response, and wrote the extracted fields to a DynamoDB table. 
Thus, each receipt was processed using Textract’s synchronous \texttt{AnalyzeExpenseCommand} API~\cite{AWSTextract3}, which returned \texttt{ExpenseDocuments} containing \texttt{SummaryFields} and \texttt{LineItemGroups}. 
\texttt{SummaryFields} would contain the vendor name, date and total, and \texttt{LineItemGroups} would contain an array of purchase items, where each array element would contain the name, price and quantity of the purchase item.
After all receipts had completed processing, the team manually exported the DynamoDB records to a CSV file for detailed analysis. This CSV export was performed only for the evaluation and is not part of the ReceiptCAT production workflow. 

Each record in the CSV included the following fields:
\begin{enumerate}
\item   Receipt\_ID (UUID generated in Lambda);
\item  User ID;
\item Vendor (extracted from \texttt{SummaryFields} of the \texttt{AnalyzeExpense} API response);
\item Date (extracted from \texttt{SummaryFields} of the \texttt{AnalyzeExpense} API response);
\item Total (extracted from \texttt{SummaryFields} of the \texttt{AnalyzeExpense} API response);
\item Items (list of item name, price and quantity extracted from \texttt{LineItemGroups} of the
\texttt{AnalyzeExpense} API response);
\item \texttt{S3\_Path}: File path of the receipt uploaded to the S3 bucket;
\item \texttt{Processed\_At}: Timestamp of receipt processing completion, which is recorded as a Unix epoch timestamp in milliseconds (i.e., the number of milliseconds elapsed since 1 January 1970,
00:00:00 UTC).
\end{enumerate}
 
The extracted CSV data were manually reviewed by the first and the second authors, who independently analysed a portion of the dataset, comparing the extracted fields with the original receipt images and noting recurring issues or anomalies. The review focused on identifying common extraction patterns related to field presence, image quality, text layout, and language variation.

%==============================
\section{{Case studies: Results}} 
\label{sec:results}

 In this section, we provide a summary of the results and our analysis.  
 The dataset of 118 receipts was manually reviewed to enable quantitative insights from the recorded observations. The manual ground-truth analysis compared each receipt image with its extracted Textract output to identify recurring extraction behaviours and issues, such as missing vendor names, inconsistent date formats, or incomplete item lists.

%-----------------------------------------------------
\subsection{Variability of the vendor name format}

Textract returned multiple string variants for the same vendor (e.g., `WOOLWORTHS GROUP
LIMITED', `Woolworths', `Woolworths Online'), but these variations arise from
differences in layout across stores and receipt types. 
Textract achieved similar accuracy for both text-based (68.7\%) and logo-based (66.7\%) receipts as shown in Table~\ref{tab:vendor_name_accuracy}. Receipts with logo-based presentation of the vendor's name were slightly more prone to inclusion of additional characters due to line breaks and partial logo segmentation. %  
Please note that receipts where the printed vendor name was recorded as `Unknown' (12 of 118) were excluded from accuracy calculations because no ground truth vendor text was available on the receipt for comparison, i.e. the results presented in Table~\ref{tab:vendor_name_accuracy} are based on the remaining 106 receipts.

\begin{table}[ht]
\centering
\caption{Accuracy of vendor name extraction by format of the vendor's name}
\label{tab:vendor_name_accuracy}
\renewcommand{\arraystretch}{1.2}
\begin{tabular}{lccccc}
\toprule
\textbf{Format } &
\textbf{Total } &
\textbf{Exact } &
\textbf{Mismatches} &
\textbf{Extracted as  } &
\textbf{Accuracy (\%)} 
\\
 &
\textbf{number} &
\textbf{Matches} &
 &
\textbf{``Unknown''} &
  \\
\midrule
Text & 67 & 46 & 19 & 2 & 68.7 \\
Logo & 39 & 26 & 13 & 0 & 66.7 \\
\bottomrule
\end{tabular}
\end{table}

%-----------------------------------------------------
\subsection{Variability of the date format}

Dates were detected consistently but appeared in varying formats such as 16 Apr 2025, 07/04/25, and2025-09-12 %(Refer to Finding 3 of the textract-analysis_appendix.pdf file within the
%P000340SE-EvaluationReport-Appendix.zip archive for date extraction samples.) 
Without post-processing, these inconsistencies might affect sorting and temporal grouping.

%-----------------------------------------------------
\subsection{Misidentification of the vendor}

In the cases when vendor names were absent, Textract sometimes misclassified nearby text as vendor names, including personal identifiers. 
If no recognisable vendor text existed, `Unknown' was returned. %(Refer to Finding 4 of the textract-analysis_appendix.pdf file within the P000340SE- EvaluationReport-Appendix.zip archive for vendor misidentification examples.)

%-----------------------------------------------------
\subsection{Language limitations}

\begin{figure}
    \centering
    \includegraphics[width=0.5\linewidth]{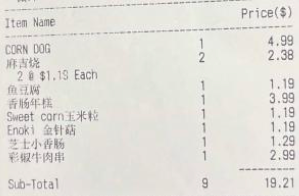}
    \caption{Extract of a receipt presenting the list of purchased items}
    \label{fig:Finding6}
\end{figure}

Receipts containing non-English product names were partially parsed. Unsupported scripts led to
missing item data, confirming that Textract’s model primarily targets English text.  Figure~\ref{fig:Finding6} provides an example: an extract of a receipt presenting the list of purchased items, where the majority of the items were specified not in English. %(Refer to Finding 6
The items-field produced by Textract for this case was as presented below:\\
\texttt{"[{""name"":""CORN DOG"",""price"":""4.99"",""quantity"":""1""},}\\
\texttt{{""name"":""Sweet corn"",""price"":""1.19"",""quantity"":""1""},}\\
\texttt{{""name"":""Enoki"",""price"":""1.19"",""quantity"":""1""}]"}

%-----------------------------------------------------
\subsection{Sensitivity to the image quality}

Image resolution and clarity directly affected extraction completeness. 
Blurry, angled, or shadowed receipts often resulted in missing or partially extracted fields. Textract also occasionally omitted data from otherwise legible images, indicating sensitivity to lighting and image noise. However, the current dataset was limited and did not include receipts captured under systematically varied lighting conditions or across multiple versions of the same receipt (e.g., crisp versus wrinkled or folded copies). 
Only a few examples of these issues were observed, and further analysis with a broader
dataset is required in the future quantitative evaluation to validate this finding. 
%(Refer to Finding 7 

%-----------------------------------------------------
\subsection{Invoice-style receipts}

\begin{figure}
    \centering
    \includegraphics[width=0.5\linewidth]{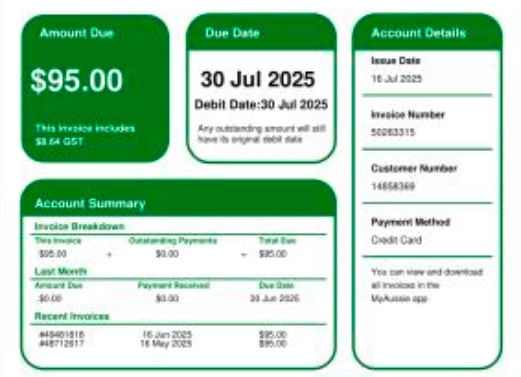}
    \caption{Extract of an invoice-style receipt}
    \label{fig:Finding8}
\end{figure}

Receipts structured as invoices lacked line-item data. Textract still extracted summary fields (vendor, total, date) but left item arrays empty, aligning with its design focus on expense-style layouts. 
Figure~\ref{fig:Finding8} provides an example: an extract of an `Aussie Broadband' receipt presenting the list of purchased items (Internet services), where the receipt has the style of an invoice. The items-field produced by Textract for this case was an empty list ([]).

%-----------------------------------------------------
\subsection{Summary of the analysis}

Based on the analysis of the results, we propose the following mitigation strategies:

\textbf{Recommendation 1.} Apply post-processing logic to standardise extracted dates and price fields. This includes converting dates to a consistent format (YYYY-MM-DD) and ensuring total price values are properly formatted as numeric data.

\textbf{Recommendation  2.} When required fields such as vendor, date, or total are missing, populate them with default values before passing the data downstream. This approach ensures continuity of processing while maintaining placeholders for later correction.

%========================================
\section{{Limitations and Threats to validity}}
\label{sec:limitations}

Our experiments face several threats to validity. 
This study was designed as an initial qualitative assessment based on manual review. % 
The dataset was collected by the authors and therefore lacks the diversity and structured sampling needed for statistical diversity across vendors, receipt types, and image conditions. Precision, recall, and F1 scores were not computed because the analysis focused on observing recurring extraction issues through direct comparisons between images and outputs. Additionally, the evaluation did not include controlled testing across variables such as lighting, angle, or receipt condition, and only the synchronous \texttt{AnalyzeExpense} API was tested. %
However, even with these limitations, the findings provide valuable insights for shaping an in-depth study and developing an early prototype. The results revealed the need for post-processing steps to normalise extracted data (such as date formats and total values) and highlighted cases in which Textract inconsistently detected vendor names, dates, or line items. 
% 

%======================================== 
\section{{Conclusions and Future Work}}
\label{sec:conclusions}

This study provided an initial, qualitative understanding of AWS Textract’s capabilities when
applied to receipt images of varied formats and conditions. 
Textract consistently identified totals across most receipts, demonstrating its potential to extract key expense information without requiring custom model training. 
However, the evaluation also revealed recurring challenges, including inconsistent vendor detection and missing or irregular line-item and date data.
Although quantitative accuracy metrics were not computed, the findings offered valuable insight into how Textract behaves under real-world conditions. The evaluation demonstrated that while Textract’s raw outputs are not immediately production-ready, they can form a
solid foundation when complemented with appropriate post-processing and validation logic.

%===================================
%\vfill
\section*{{Acknowledgements}}

We would like to thank Shine Solutions for sponsoring this project under the research grant PRJ00002626, and especially Branko Minic and Adrian Zielonka for sharing their industry-based expertise and advice.

%===================================
 
%\bibliography{lit}
 

\end{document}